\renewcommand{\bibliographystyle}[1]{}

\documentclass[%
superscriptaddress,
twocolumn,
10pt,
 amsmath,amssymb,
 aps,
 pra,
floatfix]{revtex4-1}

\usepackage{graphicx,array}
\usepackage{dcolumn}
\usepackage{multirow}

\begin{document}
%%%%%%%%%%%%%%%%%%%%%%%%%%%%%%%%%%%%%%%%%%%%
\title{Linking engagement and performance: The social network analysis perspective}

\author{Eric A. Williams}
\affiliation{Department of Physics, Florida International University, Miami, Florida 33199, USA}

\author{Justyna P. Zwolak}
\email{jpzwolak@nist.gov\newline Present address: National Institute of Standards and Technology, Gaithersburg, MD 20899, USA}
\affiliation{Joint Center for Quantum Information and Computer Science, NIST/University of Maryland, College Park, Maryland 20742, USA}
\affiliation{Department of Teaching and Learning, Florida International University, Miami, Florida 33199, USA}

\author{Remy Dou}
\affiliation{Department of Teaching and Learning, Florida International University, Miami, Florida 33199, USA}
\affiliation{Department of Physics, University of Maryland, College Park, Maryland, 20742, USA}

\author{Eric Brewe}
\affiliation{Drexel University, Department of Physics, Philadelphia, Pennsylvania 19104, USA}
\affiliation{Drexel University, School of Education, Philadelphia, Pennsylvania 19104, USA}

\date{\today}%
%%%%%%%%%%%%%%%%%%%%%%%%%%%%%%%%%%%%%%%%%%%%%%%%%%%%%%%%%%%%%
\begin{abstract}
Theories developed by Tinto and Nora identify academic performance, learning gains, and involvement in learning communities as significant facets of student engagement that, in turn, support student persistence. Collaborative learning environments, such as those employed in the Modeling Instruction introductory physics course, provide structure for student engagement by encouraging peer-to-peer interactions. Because of the inherently social nature of collaborative learning, we examine student interactions in the classroom using network analysis. We use centrality---a family of measures that quantify how connected or ``central'' a particular student is within the classroom network---to study student engagement longitudinally. Bootstrapped linear regression modeling shows that students' centrality predicts future academic performance over and above prior GPA for three out of four centrality measures tested. In particular, we find that closeness centrality explains $28\,\%$ more of the variance than prior GPA alone. These results confirm that student engagement in the classroom is critical to supporting academic performance. Furthermore, we find that this relationship for  social interactions does not emerge until the second half of the semester, suggesting that classroom community develops over time in a meaningful way.
\end{abstract}
\maketitle

%%%%%%%%%%%%%%%%%%%%%%%%%%%%%%%%%%%%%%%%%%%%%%%%%%%%%%%%%%%%%
%---INTRODUCTION---%
%%%%%%%%%%%%%%%%%%%%%%%%%%%%%%%%%%%%%%%%%%%%%%%%%%%%%%%%%%%%%
\section{Introduction}\label{sec:introduction}
%%%%%%%%%%%%%%%%%%%%%%%%%%%%%%%%%%%%%%%%%%%%
It has long been recognized that to advance our economy and our society we need to develop a strong workforce of experts in science, technology, engineering, and math (STEM)~\cite{NSF96-STF, Adkins12-ANS, PCAST12-ETE}. Yet, of all students who enter a four-year college intending to major in a STEM field, over $30\,\%$ fail to graduate with a STEM degree~\cite{NSB18-SEI}. The situation for students from historically underrepresented groups in STEM is even more alarming: of all the STEM bachelor degrees awarded nationwide, only $12.8\,\%$ goes to Hispanic students, $8.7\,\%$ to Black or African American students, and $0.5\,\%$ to American Indian or Alaska Native~\cite{NSB18-SEI}. These percentages are not commensurate with the demographic distribution of the U.S. national population. One pathway to address this issue and move toward equity in STEM is by making systemic changes to classes and departments that promote the retention and persistence of students from minority groups. 

Tinto's model of student integration proposed in the mid 1970s links both retention (the successful completion of a course) and persistence (the successful completion of a sequence of courses) to student engagement~\cite{Tinto75-DHE, Tinto97-CAC, Tinto04-SRG, Tinto06-RPWN}. More recently, it has been suggested that persistence, engagement, and performance are all interlinked and that active learning offers key advantages over traditional lecture in all these domains~\cite{Freeman14-PNAS}. The Modeling Instruction (MI) program at Florida International University (FIU) is an example of a teaching approach that strongly emphasizes active participation and hands-on learning. From the instructional design (``learning by doing'') to the flow of the activities (small group work, multigroup discussions, a whole class wrap up) to the flexible classroom space (tables arranged to allow moving about freely), everything in MI is designed to promote development of a community of learners that each student can be an integral part of~\cite{Hestenes1987, Brewe08-MTA}. Indeed, prior research shows that students in MI experience superior outcomes in learning gains, passing rates, attitudes toward science, and retention rates compared to their lecture-based counterparts~\cite{Brewe10-equity, Sawtelle10-PIMISE, Brewe13-CLASS}. 

With the community structure as its defining feature, the MI classroom offers fertile ground for examining student engagement. In a general sense, engagement is defined as a multifaceted construct that describes the behavioral, emotional, and cognitive ways in which students immerse themselves into the academic system~\cite{Fredricks04-SE}. Over the years, its meaning has grown and become more nuanced, with studies using ``integration'' and ``involvement'' interchangeably with ``engagement''---terms referring to related but arguably separate constructs---exacerbating the complexity around this term. Throughout this work, ``engagement'' is used to signify the extent to which individuals actively connect to the academic and social fabric of a learning institution~\cite{Tinto97-CAC}. This working definition allows to explore engagement at different levels within the institution (e.g., the classroom) and provides a broad array of options for defining ``connected'' (e.g., belonging to the same group or enrolling in a common course). Decades of work show that engagement is critically important to students, especially during their first year of college~\cite{Tinto06-RPWN}.

To bring some much-needed specificity to the idea of student engagement, we turn to the toolkit offered by social network analysis (SNA)~\cite{Wasserman94,Prell13-SNA}. In particular, we operationalize engagement by quantitatively mapping students and their interactions with one another onto a network and then using SNA to analyze the network structure and properties. SNA methods help reveal student interaction patterns, characterize the roles played by specific students, and identify preferential positions in the network based on, e.g., ``access'' to other people, information or resources. It is, thus, very well suited to capture aspects of engagement related to student connectedness through peer interactions.

Depending on the specific type of interactions one is examining, connectedness can take on several meanings. In the language of SNA this is captured through measures of ``centrality'' -- a family of mathematical algorithms that calculate the various positionings of individuals in a network~\cite{Wasserman94,Prell13-SNA}. It is important to stress that centrality alone does not necessarily capture the quality of connections between two people (e.g., the level of loyalty between friends) nor does it provide insights into how or why the interaction occurred. Rather, it quantifies the structural connectedness of individuals in the network. The specific parameters around what is defined as an interaction (e.g., attending the same lecture or working on a joint project) may further shed light on the quality or characteristics of the connections. 

SNA has been used before to link students' network characteristics to a variety of education-related outcomes, including concept inventory scores~\cite{Brewe16-MAMCR}, academic performance~\cite{Rizzuto09-NWW, Gasevic13-GPA, Hommes12-SNA, Vargas18-CAP, SmithPeterson07, BruunBrewe13-FCI}, persistence~\cite{Thomas00-TTB, Eckles12-RAD, Zwolak17-NIP, Zwolak18-PIN}, self-efficacy~\cite{Dou16-BPM}, and anxiety~\cite{Dou19-PGN}. Yet, no study known to us performed a longitudinal analysis of interaction data to show how engagement changes throughout the semester and how during this time the relationship between students' engagement, course grades, and past academic performance evolves. 

Our work complements previous findings connecting SNA and performance in a number of important ways. In particular, in Ref.~\cite{Vargas18-CAP}, the authors study students' homework assistance networks, but they do so on a rather global scale by looking at a single network of aggregated data without taking into account the nuanced development of interactions throughout the semester. Moreover, they study upper-division courses where students have had many opportunities to get to know each other from their lower-level courses and they ``have already had much of their undergraduate career to develop collaboration strategies that they believe work for them''~\cite{Vargas18-CAP}. In Ref.~\cite{Hommes12-SNA}, past performance is incorporated in a study of how informal social interactions influence student learning (as measured by exam performance). Yet, in their analysis the authors rely on a single data collection and do not account for the in-class interaction. Moreover, the population in this study comprised of students enrolled in a highly competitive medical school. As a postsecondary minority-serving institution, with $76.5\,\%$ of the student population coming from historically underrepresented minority groups~\cite{FIU-demog}, FIU is an important case for studying the effects of building communities on performance.

This paper is organized as follows: We begin with a review of the literature in Section \ref{sec:lit_rev}, followed by a discussion of the theoretical framework in Section \ref{sec:framework}. In Section \ref{sec:method}, we describe the methodology used in this study. We then present our results with relevant interpretation and discussion of our findings in Section \ref{sec:Results}. Finally, we close by drawing conclusions and suggesting future lines of inquiry in Section \ref{sec:conclusion}.

%%%%%%%%%%%%%%%%%%%%%%%%%%%%%%%%%%%%%%%%%%%%%%%%%%%%%%%%%%%%%%%
%---THE MAIN ARTICLE---%
%%%%%%%%%%%%%%%%%%%%%%%%%%%%%%%%%%%%%%%%%%%%%%%%%%%%%%%%%%%%%%
\section{Literature Review}\label{sec:lit_rev}
%%%%%%%%%%%%%%%%%%%%%%%%%%%%%%%%%%%%%%%%%%%%%%%%%%%%%%%%%%%%%%
Despite being well studied, student engagement is challenging to define succinctly. Its rich, nuanced, and subtle character often takes on a specific meaning from the context in which it is viewed. As mentioned earlier, throughout this work, engagement refers to the presence (or absence) of social interactions, both formal and informal, occurring inside or outside the classroom, with both peers and faculty. It does not describe these social activities qualitatively, i.e., it does not answer the question of why or in what ways students are engaging while interacting. 

In general, engagement may be self-initiated or a result of membership in some kind of group, either university-sponsored or not, and is often discussed in relation to persistence. As Tinto writes, ``though we have a sense of why involvement or integration [engagement] should matter (e.g., that it comes to shape individual commitments), we have yet to explore the critical linkages between involvement [engagement] in classrooms, student learning, and persistence''~\cite{Tinto97-CAC}. While several researchers have stepped forward to answer this call \cite{Rizzuto09-NWW, Gasevic13-GPA, Hommes12-SNA, Vargas18-CAP, SmithPeterson07, BruunBrewe13-FCI, Thomas00-TTB, Eckles12-RAD, Zwolak17-NIP, Zwolak18-PIN} there is still much work to be done; this study is a step toward filling this gap. 

%%%%%%%%%%%%%%%%%%%%%%%%%%%%%%%%%%%%%%%%%%%%%%%%%%%%%%%%%%%%%%
\subsection{Persistence and performance}
%%%%%%%%%%%%%%%%%%%%%%%%%%%%%%%%%%%%%%%%%%%%%%
Although studies of engagement have historically focused on persistence, academic performance is a common theme that permeates much of the literature. Tinto's Model of Student Integration provides a first heuristic and theoretical framework for understanding how these three constructs interlink~\cite{Tinto75-DHE}. In his seminal work, Tinto identified a critical two-step linkage ``between involvement and learning, on one hand, and between learning and persistence, on the other''~\cite{Tinto97-CAC}. After all, persisting in a program of study is impossible without some successful academic performance along the way. 

Nora expanded Tinto's model to incorporate additional components, such as individual pull factors acting as barriers to student engagement~\cite{Nora03-AHH}. While focusing on traditionally underrepresented students, he found academic performance to be ``possibly the most influential factor'' on Hispanic students' persistence. Final grades seemed to influence Hispanic students' drop-out decisions three times as much as they did for nonminorities. Nora argued that the connections from students' performance to their sense of belonging and their perception of their ability to earn a degree are essential factors when deciding whether to remain in college~\cite{Nora03-AHH}. 

In his model of student involvement, Astin went even further and reframed the decision to persist or drop out as polar ends of an ``involvement spectrum''~\cite{Astin84-SIM}. Under this new framing, dropout represents the most extreme form of disengagement at the low end of the spectrum while the rest of the spectrum corresponds to the many possible gradations of involvement, including consistent active engagement that supports a student's successful persistence. Astin categorizes involvement into different types, including residing on campus, academic involvement, interactions with faculty, and extracurricular activities, such as student government, honors programs, and athletics. 

\begin{figure}[t]
\includegraphics[width=0.48\textwidth]{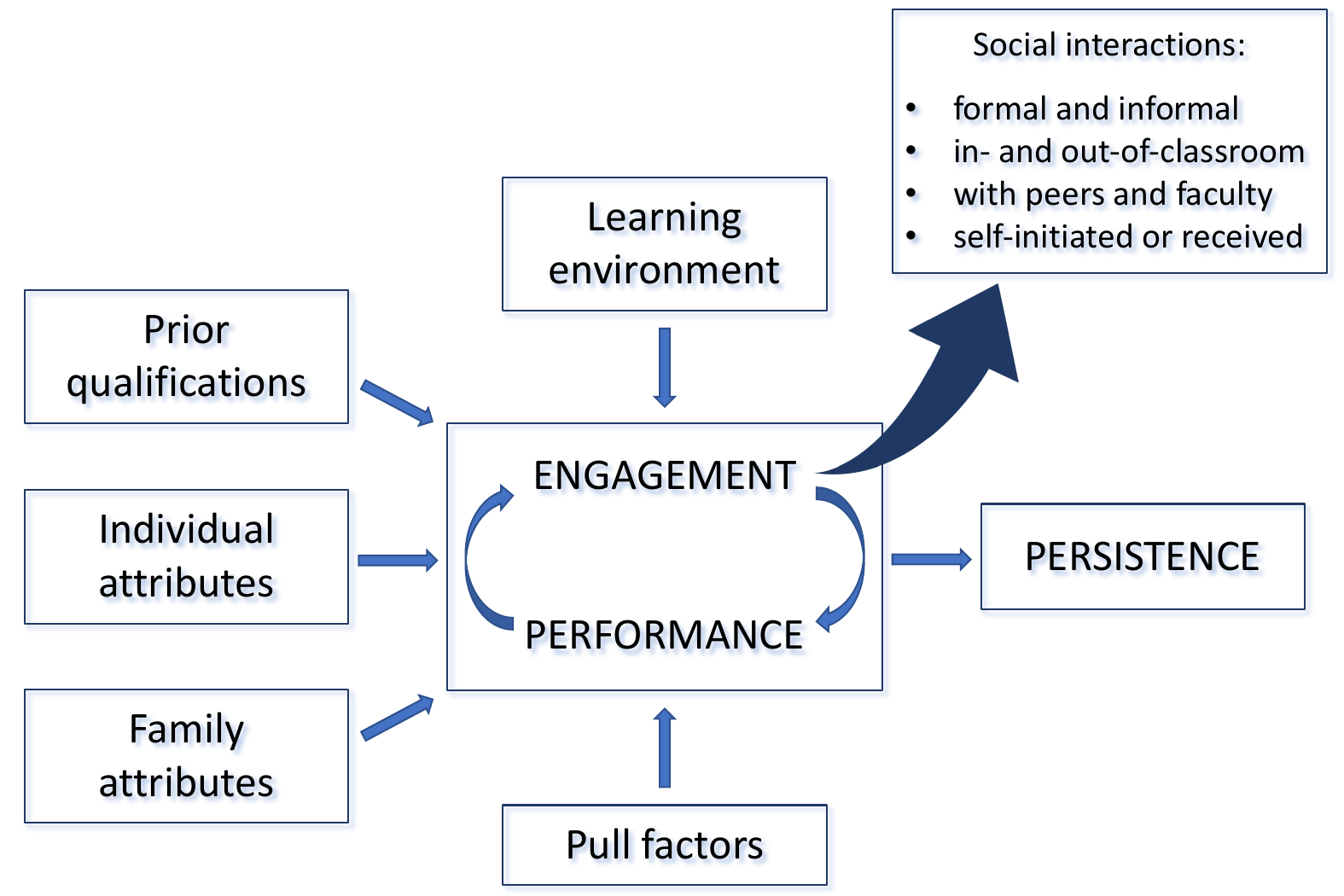}
\caption{A simplified conceptual schema of the theoretical framework combining elements from Tinto's and Nora's models used in this work~\cite{Tinto75-DHE, Nora03-AHH}.}
\label{fig:model_vis}
\end{figure}

All these models include performance as an explicit factor, providing us with the theoretical grounding for exploring a relationship between engagement and performance. The study reported in this paper is a part of a larger project that explores all three of these variables, though it does not investigate persistence explicitly. Instead, we conceive of engagement and performance as precursors to persistence (see Fig.~\ref{fig:model_vis}). Our theoretical framework, described in detail in Sec.~\ref{sec:framework}, further elaborates on this relationship.

%%%%%%%%%%%%%%%%%%%%%%%%%%%%%%%%%%%%%%%%%%%%%%%%%%%%%%%%%%%
\subsection{Network analysis in education research}
%%%%%%%%%%%%%%%%%%%%%%%%%%%%%%%%%%%%%%%%%%%%%%
Since students' interactions are inherently relational, it is natural to operationalize them with social network analysis. In fact, Tinto explicitly called for the use of SNA, saying ``we would be well served ... to study the process of persistence with network analysis and/or social mapping of student interaction patterns''~\cite{Tinto97-CAC}. Thomas answered the call posed by Tinto using SNA to examine a nuanced relationship between self-reported interactions and multiple factors in Tinto's model, including GPA and persistence into subsequent courses~\cite{Thomas00-TTB}. Since then, education researchers have linked network measures to a variety of constructs found in the engagement literature, including academic performance~\cite{Rizzuto09-NWW, Gasevic13-GPA, Hommes12-SNA, Vargas18-CAP, SmithPeterson07, BruunBrewe13-FCI}, persistence~\cite{Thomas00-TTB, Eckles12-RAD, Zwolak17-NIP, Zwolak18-PIN}, self-efficacy~\cite{Dou16-BPM}, anxiety~\cite{Dou19-PGN}, and sense of community~\cite{Dawson08-RSC}. Yuan {\it et al.} and Gasevic {\it et al.} used network measures to define social capital in distance-learning courses. Using four different centrality measures, they examined the development of students' social capital over time (in multiple courses) and its impact on academic performance (course grades and cumulative GPA)~\cite{Yuan06-comm, Gasevic13-GPA}. Mayer and Puller used SNA on Facebook ``friends'' data to investigate the formation of cliques at ten public universities in Texas. In their analysis, they consider both environmental factors of each school and personal attributes of the students~\cite{MayerPuller08-old}. Forsman {\it et al.} used network analysis to interpret existing persistence literature and describe how ``the networked interactions, the social system, and the academic system are all coadapting'' over time~\cite{Forsman14-CSN}. 

Including students' network centrality scores as an explicit term in a structural equation model led Hommes {\it et al.} to two key results relevant to our study: (i) centrality was predicted by high school GPA, and (ii) centrality predicted future performance on a multiple-choice ``factual knowledge test''~\cite{Hommes12-SNA}. It is notable that the authors computed centrality from three types of out-of-class networks (friendship, giving information, receiving information). They found in all cases that centrality was a significant predictor even when controlling for age, gender, high school GPA, academic motivation, and institutional-level social integration. However, this study was limited by its discussion of only one type of centrality and did not take into account content-related interaction. Furthermore, the student population was exclusively high-GPA Dutch medical students in the Netherlands; results from such a setting do not necessarily generalize to minority students in the United States. 

Bruun and Brewe found that course grades were predicted by network centrality scores after controlling for students' FCI prescores~\cite{BruunBrewe13-FCI}. Interestingly, they found that the most-predictive centrality scores came from a primarily social communication network, not from the two networks created based on content-related communication. Such a result defies what one may expect---that information flow about the subject matter would be more relevant to performance than socializing. It also differs from an earlier report by Smith and Peterson who found quiz and paper grades to be positively correlated with peer-reported interactions about course advice, but negatively correlated with peer-reported interactions about general advice~\cite{SmithPeterson07}. These two studies vary in several key ways (student population, centrality measures calculated, performance metrics used) and thus are not directly comparable. However, they highlight the powerful impact of the classroom on students' overall academic experience, which motivates further study of the in-class peer interactions.

%%%%%%%%%%%%%%%%%%%%%%%%%%%%%%%%%%%%%%%%%%%%%%%%%%%%%%%%%%%%%%
\section{Theoretical framework}\label{sec:framework}
%%%%%%%%%%%%%%%%%%%%%%%%%%%%%%%%%%%%%%%%%%%%%%%%%%%%%%%%%%%%%%
Our theoretical framework, depicted in Fig.~\ref{fig:model_vis}, is based mainly on Tinto's model of student integration, though we also draw from Nora's work on student engagement and Astin's theory of involvement~\cite{Tinto75-DHE, Nora03-AHH, Astin84-SIM}. %, Note1}. 
We apply these theories to build a coherent scaffold for understanding students' immersion into the social and academic spheres of their learning community, and the effects this immersion has on them. 

%%%%%%%%%%%%%%%%%%%%%%%%%%%%%%%%%%%%%%%%%%%%%%%%%%%%%%%%%%%%%%
\subsection{The nature of engagement}\label{subsec:nat_of_engag}
%%%%%%%%%%%%%%%%%%%%%%%%%%%%%%%%%%%%%%%%%%%%%%
Students' peer interactions may occur in a variety of ways. They may take place in the classroom or outside of it; they may be related to course content, extracurricular activities or personal life; they may occur with other students or with faculty; they may occur in settings formal or informal. All of these types of connections, and more, contribute to a student's integration within the social and academic fabric of the institution. A large body of research on student engagement depicts a multidimensional understanding of what engagement means in varying contexts~\cite{Fredricks04-SE}. 

Of all the possible places for student engagement to develop, the college classroom is perhaps the most important. As Tinto writes, ``It is evident that participation in a collaborative or shared learning group enables students to develop a network of support ... engaging them more fully in the academic life of the institution'' with the classroom learning community becoming ``a gateway for subsequent student involvement'' within the institution~\cite{Tinto97-CAC}. This is especially true for first-year students, who have not yet established a support network, and commuters, who must attend to a variety of off-campus responsibilities throughout the day. Thus, we focus exclusively on student-to-student interactions occurring in the classroom and during class time, drawing from the above mentioned models to guide our investigation of student engagement as it manifests in our specific context. In particular, from Tinto, we focus on supportive, informal peer group associations; perceptions of ``social fit''; bridging the academic-social divide; and gaining a voice in the construction of knowledge. From Nora, we highlight in-class experiences and collaborative learning as ways to be part of a learning community; peer group interactions as a meaningful social experience; and academic performance as an important cognitive outcome. Behaviors supporting all of these mechanisms are cultivated by the MI curriculum, and are further encouraged by the course instructors.

It is important to stress that engagement does not always come easy. There exist many barriers (inside and outside the classroom) that hinder meaningful student engagement. Among these, Nora identifies \textit{pull factors}, both tangible and intangible, that serve to drag students out of the community. These include family and work responsibilities, the need to commute to school, and financial needs. For underrepresented groups, including minority students, the presence or perception of prejudice and discrimination on campus is especially harmful~\cite{Nora03-AHH}. Pull factors are everyday realities for FIU students---the vast majority of whom commute and come from working class families; over half of whom are first-generation college students \cite{FIU-housing, FIU_OPIR16-AAP, First-Gen}---and the specific students in our classroom, $85\,\%$ of whom are minority students.

%%%%%%%%%%%%%%%%%%%%%%%%%%%%%%%%%%%%%%%%%%%%%%%%%%%%%%%%%%%%%%
\subsection{The relationship between engagement and performance}
%%%%%%%%%%%%%%%%%%%%%%%%%%%%%%%%%%%%%%%%%%%%%%
Based on a synthesis of the literature, our framework leads us to hypothesize a direct positive relationship between student engagement and academic performance. Since learning in a MI classroom happens through social interactions, there is a case to be made that more frequent and more effective engagement corresponds to better learning. Consequently, a better learning experience should lead to better academic performance (and, ultimately, persistence). Tinto identified performance as a critical intermediate link in a two-stage relationship between engagement and learning on one side, and learning and persistence on the other~\cite{Tinto97-CAC}. Thus, we have sufficient theoretical grounding to expect that engagement contributes to academic performance.

However, we are aware that the relationship may point the other way~\cite{MayerPuller08-old}. For example, a student who performs well on an exam may receive a confidence boost that leads them to speak more freely in class discussions, or be sought out by other students as a study partner. Performing poorly on an exam, on the other hand, may discourage a student from participating actively in the future. These are but a few examples of how past performance may influence future engagement; the literature reinforces this idea. Nora's model explicitly includes GPA and cognitive gains (both perceived and actual) as factors contributing to student engagement~\cite{Nora03-AHH}. 

We propose that both directions of this influential pathway are possible. We claim that student engagement and academic performance exhibit a relationship that is reciprocal and iterative, i.e., that past performance influences engagement which in turn influences future performance~\cite{Hommes12-SNA, Williams15-UCN}. Therefore, we hypothesize that student engagement in the MI classroom will predict future academic performance even when controlling for prior qualifications.

%%%%%%%%%%%%%%%%%%%%%%%%%%%%%%%%%%%%%%%%%%%%%%%%%%%%%%%%%%%%%%
\subsection{Formation of social networks}
%%%%%%%%%%%%%%%%%%%%%%%%%%%%%%%%%%%%%%%%%%%%%%
As we are interested in the student-to-student social interactions occurring within a MI classroom, it is natural to use a relational data analysis tool. SNA allows one to study the network representation of the in-class interactions in many ways and at many levels: from visualizing the entire network, to examining its overall structure and cohesion, to quantifying individuals' embeddedness within the network (both globally and at a local, ``nearest neighbors'' level). Using the various measures of network embeddedness (i.e., the centrality scores) as a proxy for engagement, we can explore the relational position of a given student with respect to the rest of the class, and to quantify that student's engagement within the classroom community.

Cross-group interactions are inherent to the design of the MI course. They are explicitly encouraged during instructor-moderated ``board meetings'' when multiple groups come together to discuss their work. Since groups frequently present solutions to different problems featuring various aspects of a physical phenomena, intergroup discussion is a necessity for each group to understand what the other groups did, how they did it, and why. In addition, unstructured cross-group interactions outside of the board meetings occur on a regular basis, e.g., during problem-solving sessions and experimental investigations. Course instructors permit this cross-group talk, though after some minutes they generally encourage students to return to their assigned groups to share and implement the discussed ideas. 

Integration into a community does not, however, happen instantaneously. Rather, it is a gradual process. We expect that over time, two things will happen: students will get to know more of their peers and to know each other better. In other words, both the quantity and quality of interactions will increase. The MI curriculum is expressly designed to encourage, even require, collaborative work and the course instructors explicitly promote its benefits. At the same time, since this course structure differs from traditional physics classrooms, students might need some time to adjust but will, as time passes, engage more often and more effectively with each other. Finally, while it is natural that in-class interactions occur more often between students in close physical proximity, we expect to see this preference for interacting exclusively with peers in the same group or table significantly decrease over time. We hypothesize a progression from little interaction at first, to interaction with only seatmates, to some interaction outside of their table, leading finally to moderate interaction with many peers outside of their table.

%%%%%%%%%%%%%%%%%%%%%%%%%%%%%%%%%%%%%%%%%%%%%%%%%%%%%%%%%%%%%%
\subsection{Research objectives}
%%%%%%%%%%%%%%%%%%%%%%%%%%%%%%%%%%%%%%%%%%%%%%
With this framework in mind, we set out to answer following research questions:
\begin{itemize}
\item Does the in-class student engagement predict future academic performance, even when controlling for prior qualifications? 
\item Which centrality measure is the most informative in the context of in-class student engagement?
\item How does the in-class student community develop over time? When during the semester do student interactions become important to future academic performance? 
\end{itemize}

%%%%%%%%%%%%%%%%%%%%%%%%%%%%%%%%%%%%%%%%%%%%%%%%%%%%%%%%%%%%%%
\section{Methodology}\label{sec:method}
%%%%%%%%%%%%%%%%%%%%%%%%%%%%%%%%%%%%%%%%%%%%%%%%%%%%%%%%%%%%%%
This study is conducted at FIU, a large public research university serving (at the time of this study) about 54\,000 students in Miami, FL. The data come from one semester (16 weeks) of a MI introductory physics course. There were 73 first-year students enrolled ($32.9\,\%$ female), taught by one professor, two teaching assistants, and three learning assistants~\cite{LA-paper}. The demographics of the classroom is as follows: $74.0\,\%$ Hispanic, $11.0\,\%$ Black, $5.5\,\%$ Asian, $5.5\,\%$ White, and $4.0\,\%$ other.
% taught in the fall of 2015

%%%%%%%%%%%%%%%%%%%%%%%%%%%%%%%%%%%%%%%%%%%%%%%%%%%%%%%%%%%%%%
\subsection{Learning environment}
%%%%%%%%%%%%%%%%%%%%%%%%%%%%%%%%%%%%%%%%%%%%%%
The MI curriculum is an example of a teaching environment where students can actively participate in the learning process. In this particular classroom, students work together in small groups of three to conduct experiments and solve problems. The course instructor establishes the groups using an assigned seating arrangement. The arrangement is modified approximately every 3-4 weeks (usually after major in-class events, such as exams and lab experiments). The instructor makes these decisions for purely course-related reasons, independent of the research team. 

Two small groups are seated at the same round table and are encouraged to interact with one another while solving problems. It is worth noting that, in practice, students often interact with peers across groups and across tables. As a result, their academic engagement includes both intergroup and intragroup interactions on a regular basis, which is characteristic for this course structure. In addition, two or three times during the class time, small groups are meeting together in large board meetings of about 18 to 21 students where they can ask questions, present solutions, and discuss underlying physical phenomena. An implicit aim of the board meetings is to reach consensus through discussion based on experimental evidence without explicitly providing students the solution. The discussion is facilitated by instructors and undergraduate learning assistants. In short, students collaborate and interact with one another to construct their own understanding. 

%%%%%%%%%%%%%%%%%%%%%%%%%%%%%%%%%%%%%%%%%%%%%%%%%%%%%%%%%%%%%%
\subsection{Academic performance data}
%%%%%%%%%%%%%%%%%%%%%%%%%%%%%%%%%%%%%%%%%%%%%%
For the purpose of this study, we divide students' academic performance data into two categories: past and future. Past performance is represented by a student's GPA prior to the MI course, expressed on the typical 4.0 scale. Future performance is represented by student's final grade in the course (see Fig. S1 in the Supplemental Material~\cite{supp} for histograms of past and future performance measures). Note that the past performance is used only as a control variable capturing prior qualifications and we do not consider gains in a pre-post format.

In this particular class, only full letter grades are assigned, recorded using the standard grade point per unit conversion scheme (A=4.0, B=3.0, C=2.0, D=1.0, F=0.0). Students who drop or withdraw after the enrollment deadline are assigned a final grade of zero. The grades dataset was downloaded from FIU's electronic record system. Data management and analysis is done using the statistical programming language R~\cite{R, broom, dplyr}.

%%%%%%%%%%%%%%%%%%%%%%%%%%%%%%%%%%%%%%%%%%%%%%%%%%%%%%%%%%%%%%
\subsection{Network data collection}\label{subsec:survey}
\begin{figure}[t]
 \includegraphics[width=0.45\textwidth]{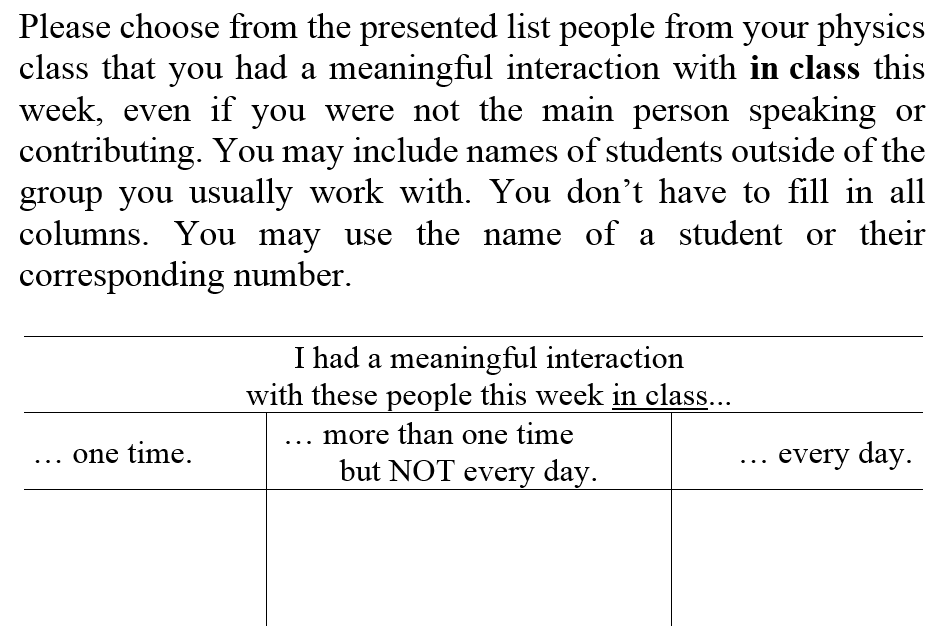}
 \caption{An excerpt from the SNA survey that was given in the Modeling Instruction classroom.~\cite{Zwolak17-NIP}}
\label{fig:SNAsp15}
\end{figure}

%%%%%%%%%%%%%%%%%%%%%%%%%%%%%%%%%%%%%%%%%%%%%%
While there are various ways to define and collect student interaction data (e.g., video analysis of students group work, discussions on a web-based social forum; see Ref.~\cite{Dou19-PGN} for additional approaches), we measure student interaction using a pen-and-paper survey, shown in Fig.~\ref{fig:SNAsp15}. Content and face validity of the survey was established through iterative discussions between the authors and other SNA experts, as well as through findings from other studies (see, e.g., Refs.~\cite{Dou16-BPM, Dou19-PGN, Zwolak17-NIP, Zwolak18-PIN}). 

The survey asks students to identify who they had meaningful in-class interactions with during a week prior to a given collection. A roster of all students enrolled in the course is included with the survey to assist students' reporting. Teaching staff is also included on the roster and appeared in the data. However, to focus exclusively on students' peer interactions we removed teaching staff from the network prior to analysis.

Each reported interaction is inherently directional: when a survey respondent marks down another student's name, the interaction is recorded as initiated by the respondent and received by the other student. Directionality indicates which student considered a particular interaction ``meaningful'' enough to remember and report on the survey and which interactions were mutual. Moreover, to explore different levels of meaningful, students are asked to indicate how often they interacted with each identified person, choosing from ``one time'', ``more than one time'', and ``every day''.

The survey was administered five times throughout the semester, spaced approximately three weeks apart (weeks 2, 6, 8, 11, and 13). Response rates were on average 83.8\,\% (unbiased SD=8.4\,\%).

\begin{figure}[t]
\centering
\includegraphics[width=1.0\linewidth]{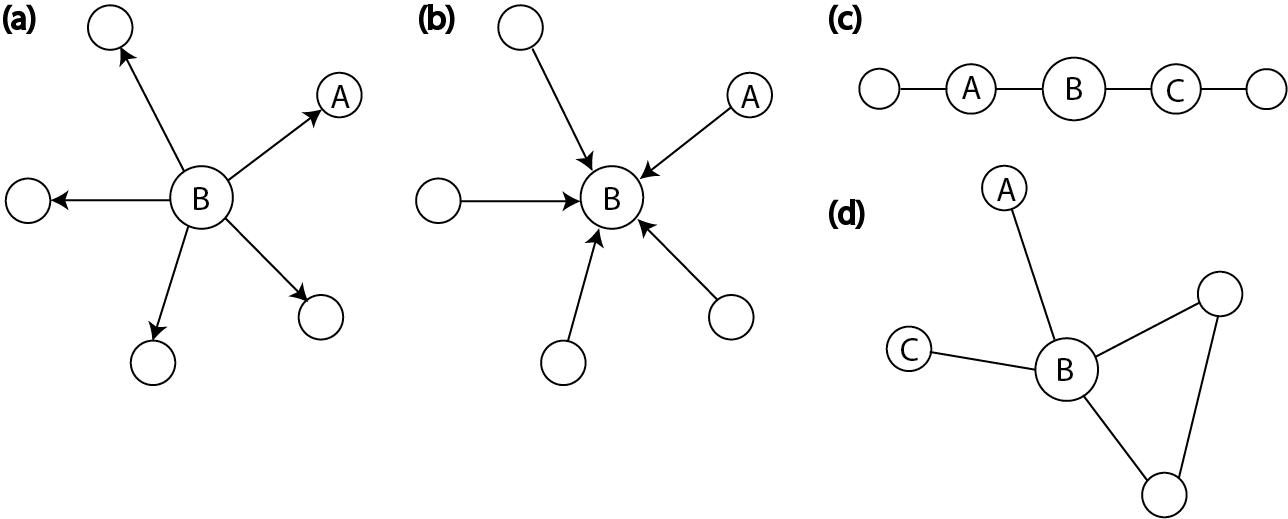}
\caption{Visualization of the four centralities used in this work. In each case, B has higher centrality than A according to (a) outdegree (a measure of engagement), (b) indegree (a measure popularity or sociability), (c) betweenness (a measure of control over the flow of information through the network), and (d) closeness (a measure of overall connectivity or embeddedness within the entire network). Adapted from Ref.~\cite{Zwolak17-NIP}.\label{fig:centralities}}
\end{figure}

%%%%%%%%%%%%%%%%%%%%%%%%%%%%%%%%%%%%%%%%%%%%%%%%%%%%%%%%%%%%%%
\subsection{Conceptualizing student engagement}\label{subsec:SNA_terminology} 
%%%%%%%%%%%%%%%%%%%%%%%%%%%%%%%%%%%%%%%%%%%%%%
Network analysis allows us to quantify classroom interactions by framing the student community in a relational way: we conceptualize the community of people as nodes and the interactions between them as ties~\cite{Prell13-SNA}. The arrangement of nodes and ties represents the social structure of the classroom community. If student A reports an interaction with student B, then the tie points from A to B. If, instead, student B reports an interaction with student A, then the tie points from B to A. If both students reported an interaction with each other, then the tie is considered bidirectional. The reported frequency of interactions is used to ascribe numerical weights to ties: ``one time'' is coded as 1, ``more than one time'' is coded as 2, and ``every day'' is coded as 3. A student's position in the network is purely relational, dependent on their ties to the other students. 

Ties are used to quantify the level of engagement within the community through \textit{centrality}. As there is more than one way to be engaged, there is an entire family of centrality measures, each accounting for the network ties in a unique way. Consequently, each centrality measure represents a different way that students integrate within the classroom community. In this study, we consider four centrality measures (see Fig.~\ref{fig:centralities}). We divide them into two categories based on the cross sections of the network structure they pertain to: \textit{local}, which include only a student's nearest neighbors, and \textit{global}, which depend on all of the students in the network~\cite{Grunspan14-SNA}. The following descriptions are intended to ground the centrality measures in a classroom context. Rigorous mathematical discussion of these measures is beyond the scope of this paper; for a more thorough treatment, see Refs.~\cite{Wasserman94, Prell13-SNA}.

For local measures, we consider only the direct connections between adjacent nodes, paying special attention to the direction of each interaction. The number of interactions reported by a given student, i.e., the number of ties pointing away from a given node, is called \textit{outdegree} [see Fig.~\ref{fig:centralities}(a)]. Since outdegree is calculated exclusively from self-reported interactions, we use it to capture a student's self-perception of their engagement in the classroom. 

We are also interested in students' engagement behavior according to their peers. The number of times a given student appeared on other students' surveys, i.e., the number of incoming ties pointing toward a given node, is called \textit{indegree} [see Fig.~\ref{fig:centralities}(b)]. Indegree can be thought of as a measure of popularity or sociability, as perceived by other students in the network~\cite{Prell13-SNA}.

The global measures are more complex and require additional framing. They describe the position of a node within the whole network. This position is strictly relational, describing where a student is located in terms of their connections to others. Not all connections are direct, though. If student A and student C report no interactions with each other, there is no tie between them. Yet, if each of them interacted with student B, then A and C have an indirect path connecting them (A to B to C) where B acts as an intermediary [see Figs.~\ref{fig:centralities}(c) and 3(d)]. In a sense, the indirect path between A and C is two steps long. This distance is defined as \textit{path length}: the number of ties separating two nodes. In a large network where two nodes may be connected by many paths, it is useful to identify the shortest path (the \textit{geodesic}). Both global measures we consider, \textit{closeness} and \textit{betweenness}, depend on the geodesic.

Betweenness centrality indicates how often a node lies on the geodesic between two other nodes [see Fig.~\ref{fig:centralities}(c)]. In the example above, B is in between A and C because B lies on the geodesic connecting them. In the classroom context, a student with high betweenness may occupy a position where they act as a bridge between two or more tightly connected groups of students -- groups that would otherwise be isolated from each other. Such a position puts the student in a gatekeeper role, giving them control of information flow within the network.

Finally, closeness centrality captures how ``close'' a node is to all other nodes in the network [see Fig.~\ref{fig:centralities}(d)]. Stated simply, one node is close to another node if they are separated by a short path length. In the example, A and B are closer together than A and C. Closeness represents how deeply a student is embedded in the community as a whole; it represents a student's ability to access others in the classroom network easily (without going through many intermediaries).

%%%%%%%%%%%%%%%%%%%%%%%%%%%%%%%%%%%%%%%%%%%%%%%%%%%%%%%%%%%%%%
\subsection{Network analysis}
%%%%%%%%%%%%%%%%%%%%%%%%%%%%%%%%%%%%%%%%%%%%%%
Data from each survey collection is converted into a list of connections (so-called ``edge list''). Interactions are weighted at three levels based on the frequency of occurrence: interactions that occurred more often were given greater weight. For network analysis, we use the igraph and tnet packages~\cite{igraph,tnet} to calculate centrality scores, using the weighted version of each centrality measure. For outdegree and indegree, we use the function \textit{strength} (i.e., the sum of weights assigned to the node's direct connections) with the direction parameter set to count only outgoing or only incoming ties, respectively. For betweenness and closeness, we utilize the built-in weight parameter offered by their corresponding tnet functions (see Table~S1 in the Supplemental Material~\cite{supp} for descriptive statistics for all centrality measures). 

It is important to note that the network survey response rate was never 100\,\%. As a result, our data sample contains missing nodes. At the same time, as we are conducting longitudinal analysis, we need a consistent list of nodes for the whole semester. While imputation may generally be used to accommodate missingness in data, it did not seem applicable in our case. Imputation ``fills in'' the missing data values without changing the pre-existing ones which is not appropriate for interdependent centrality measures. In practice, changing the centrality of a single node will affect centrality for at least one other node. Thus, we choose to address the issue with an approach that is consistent with network methods~\cite{SmithMoody13-NMR}: by carefully defining network boundaries. In particular, we create the network roster by listing all of the students who were registered for the course after the enrollment deadline. In this way, any student who appeared on the roster but did not appear in a given collection was added as an \textit{isolate} to that collection's network (an isolate is a node with no connecting ties). Importantly, since all of the data collections occurred after the enrollment deadline this method preserves any student who dropped out of the course after that date as an isolate, which is reflective of their noninvolvement, and is immune to additional students enrolling. Centralities were computed once the roster was established.

The computed centrality scores are incorporated as node attributes. Students' academic performance data (past: GPA; future: course grade) are also stored as node attributes (see the Supplemental Material~\cite{supp} for descriptive statistics of all data used in the analysis). This process ensures consistent one-to-one matching between centrality data and performance data.

%%%%%%%%%%%%%%%%%%%%%%%%%%%%%%%%%%%%%%%%%%%%%%%%%%%%%%%%%%%%%%
\subsection{Statistical analysis}
%%%%%%%%%%%%%%%%%%%%%%%%%%%%%%%%%%%%%%%%%%%%%%
As a first step, we seek to verify that course grade is predicted by precourse GPA to establish a benchmark consistent with prior work~\cite{Williams15-UCN}:
\begin{equation}\label{eq:benchmark}
M_{base}\,:\,final \; grade \sim GPA.
\end{equation}
This benchmark serves as a standard against which all subsequent models are compared. We then perform a statistical analysis on the five data collection samples. Each collection is analyzed independently as described in the following paragraphs.

To determine if any of the four centrality measures predicts future academic performance, we conduct an exploratory series of bootstrapped simple linear regressions. Linear regression modeling relies on the assumption that data is normally distributed and independent. However, this is not the case for centrality measures (see Table~S1 the Supplemental Material~\cite{supp} for descriptive statistics for centrality measures)~\cite{Albert02-SMCN}. To account for this, we use the bootstrap method~\cite{Fox15-bootstrap}. Bootstrapping is a permutation technique in which dataset values are randomly resampled to run a statistical test a large number of times. The bootstrapped statistical test results are then constructed into a distribution of values, from which a confidence interval (CI) can be calculated. If the CI excludes zero, the test result is considered statistically significant. In our analysis, we apply this technique to the results of the linear regression models: for each model of interest we run a corresponding bootstrapped linear regression with 1000 iterations to build a 95~\% CIs on the regression coefficients (estimates) and $R^2_{adj}$ values. Thus we are able to ensure the validity of our statistical results in spite of the centrality scores' interdependence and non-normality. All linear regression models discussed in this paper are bootstrapped in this way. 

The four bootstrapped simple models are of the form 
\begin{equation}\label{eq:simple}
M_{simple}\,:\,final \; grade \sim centrality \\,
\end{equation}
where $Centrality$ $\in$ \{indegree, outdegree, closeness, betweenness\}. These are used to corroborate prior results~\cite{Williams15-UCN}. 

Next, we perform a series of bootstrapped multiple linear regression models to determine whether each centrality measure's significance survives when controlling for past performance. These multiple regression models, which we will call \textit{full models}, are of the form 
\begin{equation}\label{eq:full}
M_{full}\,:\,final \; grade \sim GPA + centrality 
\end{equation}
where, again, $Centrality$ $\in$ \{indegree, outdegree, closeness, betweenness\}. As there is the possibility of collinearity between explanatory variables in multiple regression modeling, we calculate the variance inflation factor (VIF) for each collection to check for any collinearity between GPA and the centrality scores. The VIF, ranging from 1.0 to 1.13, indicates no collinearity between variables.

Finally, we use the likelihood ratio test to compare full models to the benchmark. Doing so allows us to judge whether the full models are statistically different from the benchmark, and thus select the appropriate model to explain the relationship between course grade, centrality, and prior GPA.

Because of the large number of regression tests performed, there is a concern of encountering type I error (i.e., false positive) and inferring a relationship spuriously. This is corrected by making Bonferroni adjustments to the $p$ values in order to maintain valid alpha levels (i.e., by correcting the significance level proportionally to the number of tests in a given trial). Since each survey collection is an independent dataset, the Bonferroni corrections are made at the collection level (i.e., by scaling the $p$ values by a factor of nine). Unless otherwise stated, all $p$ values reported throughout this paper are adjusted this way. We consider results with Bonferroni corrected $p<0.05$ as significant.

%%%%%%%%%%%%%%%%%%%%%%%%%%%%%%%%%%%%%%%%%%%%%%%%%%%%%%%%%%%%%%
\section{Results and Interpretation}\label{sec:Results}
%%%%%%%%%%%%%%%%%%%%%%%%%%%%%%%%%%%%%%%%%%%%%%%%%%%%%%%%%%%%%%
We first seek to establish a benchmark against which all other models can be compared. The benchmark model, given by relation~(\ref{eq:benchmark}), shows that the course grade is indeed predicted by GPA with standardized coefficient $\beta = 0.46$, standard error of estimate $SE = 0.11$, significance level $p<0.001$, $F$ statistic $F(1,71) = 18.5$, and adjusted R-squared $R^2_{adj} = 0.196$. This verifies our expectation that future performance would be predicted by prior qualification.

Next, we test the four simple regression models, given by relation~(\ref{eq:simple}), for all five network collections. Each model is tested independently of all the others. Simple models are tested to establish a proof of concept before moving on to tackling our research questions. The results, shown in Table~\ref{tab:simple_models}, are consistent with previous work~\cite{Williams15-UCN} (see Table~S3 the Supplemental Material~\cite{supp} for a table with regression results for all tested simple models). For each centrality, we report the standardized coefficient $\beta_X$ ($X\in\{I,O,C\}$, where $I$ denotes indegree, $O$ denotes outdegree, and $C$ denotes closeness), $F$ statistic $F_{1,71}$, and adjusted R-squared $R^2_{adj}$. With a statistically significant $F$ statistics, the data give evidence that the best-fitting linear model of the specified type has at least one predictor with a nonzero coefficient (i.e., that fit of the tested model is significantly better than of the intercept only model). Moreover, while R-squared provides a measure of strength of relationship between the predictors and the response variables, F-statistic allows to determine whether that relationship is statistically significant. Since betweenness is not significantly correlated with final grades when tested in the simple model, we do not include it in the analysis of full models. 

%%%%%%%%%%%%%%%%%%%%%%%%%%%%%%%%%%%%%%%%%%%%%%%%%%%%%%%%%%%%%%
\subsection{Predicting the future, accounting for the past}
%%%%%%%%%%%%%%%%%%%%%%%%%%%%%%%%%%%%%%%%%%%%%%
To answer our first research question, we build three multiple regression models that incorporate prior GPA [see Eq.~(\ref{eq:full})], with each collection network analyzed independently of any other. This corresponds to a total of 15 models with GPA as a control variable. Full models' statistics are reported in Table~\ref{tab:full_models} (see Table~S4 the Supplemental Material~\cite{supp} for a table with regression results for all tested full models). Again, we report the standardized coefficients for GPA ($\beta_{GPA}$) and the appropriate centrality ($\beta_X$ with $X\in\{I,O,C\}$), $F$ statistic ($F_{2,70}$), and adjusted R-squared ($R^2_{adj}$). We then use the likelihood ratio test (LRT) to compare each model against the benchmark to determine if they are statistically distinct. The results are shown in Table~\ref{tab:LRT}. 

\begin{table}[t]
\renewcommand{\arraystretch}{1.1}
\renewcommand{\tabcolsep}{4pt}
\caption{Linear regression results for the simple models (final grade $\sim$ centrality). Reported $p$ values have been Bonferroni adjusted at the collection level. Significant $p$ values are marked with the appropriate number of asterisks. Models in which centrality is not a significant predictor and models that failed the bootstrap test are omitted for clarity. This includes the entire week 2 and the betweenness centrality throughout the semester.} \centering
\begin{tabular}{lrcccc}
\hline \hline
Centrality &   &\multicolumn{4}{c}{Regression statistics}\\
& & Week 6 & Week 8 & Week 11 & Week 13 \\
\hline
Indegree & $\beta_I$ & - & $0.40^{**}$ & $0.55^{***}$ & $0.36^{*}$ \\
&  $F_{1,71}$ & - & 13.7 & 30.6 & 10.4 \\
&  $R^2_{adj}$ & - & 0.15 & 0.291 & 0.115 \\
\hline

Outdegree  & $\beta_O$ & $0.43^{**}$ & $0.52^{***}$ & $0.40^{**}$ & $0.44^{**}$ \\
&  $F_{1,71}$ & 16.3 &
26.2 & 13.3 & 16.7 \\
&  $R^2_{adj}$ & 0.175 &
0.259 & 0.145 & 0.179 \\
\hline

Closeness  & $\beta_C$ & - & $0.50^{***}$ & $0.55^{***}$ & $0.44^{***}$ \\
&  $F_{1,71}$ & - & 24.1 & 30.7 & 17.0 \\
& $R^2_{adj}$ & - & 0.243 & 0.292 & 0.182 \\

\hline \hline
\multicolumn{6}{l}{\footnotesize ***$p<0.001$, **$p<0.01$, *$p<0.05$}
\label{tab:simple_models}
\end{tabular}
\end{table}

\begin{figure*}
\centering
\includegraphics[width=\textwidth]{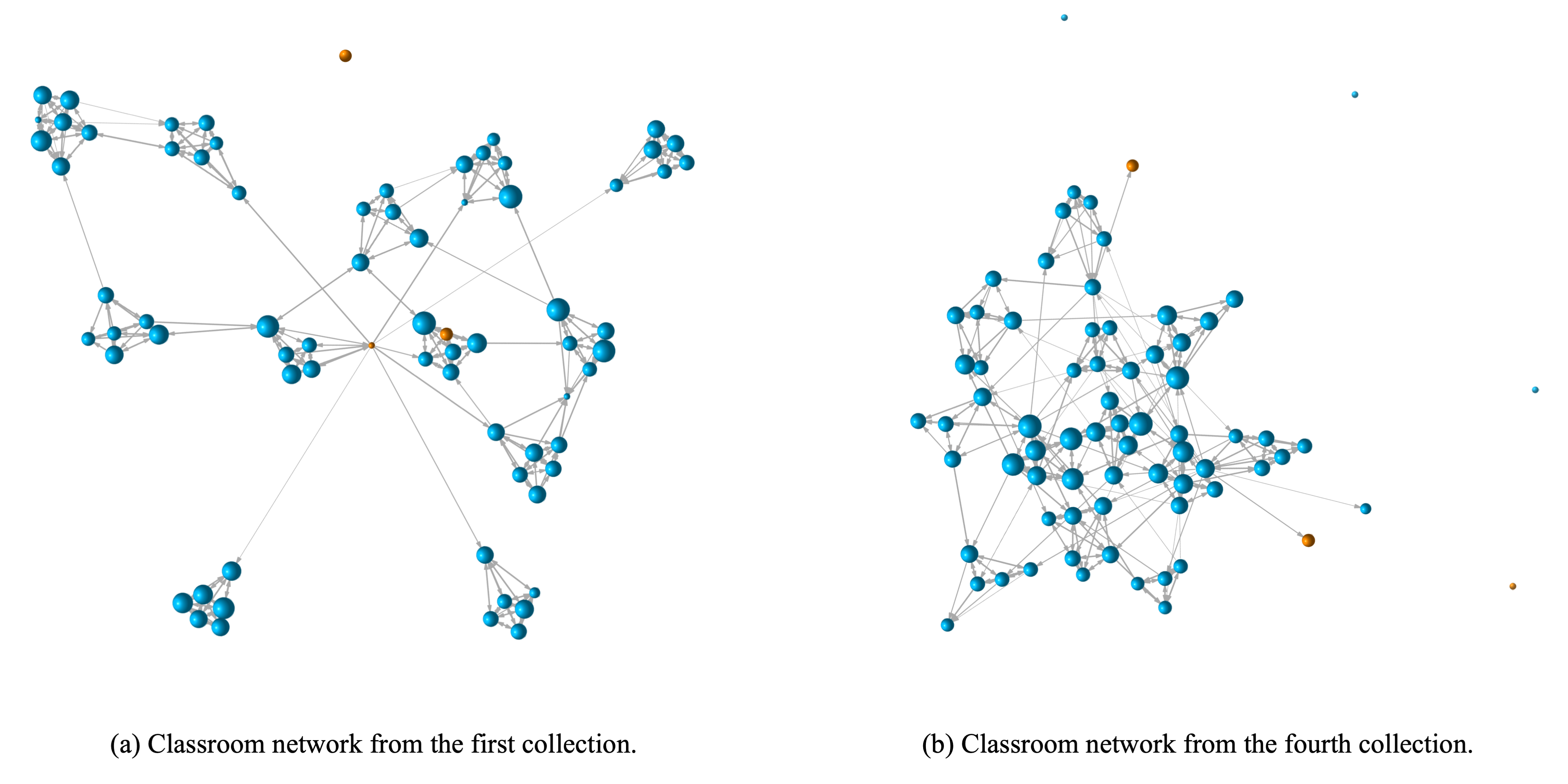}
\caption{Network visualization for two different collection times. At the first collection, ties largely represent seating arrangement; at the fourth collection, the structure of ties is demonstrably more complex. This structural difference is indicative of the network's development over time. The three orange nodes indicate students who did not complete the class. It is interesting to observe how they over time moved from fairly central position in the network (a) to being either at a peripheral position or completely disconnected by week 11. The size of nodes represents closeness at the fourth collection.} \label{fig:sna_development}
\end{figure*}

In the first collection, no centrality measure emerges as a significant predictor of final grade. Thus we conclude that integration levels at such an early time during the semester do not predict future academic performance. A comparison of the network [visualized in Fig.~\ref{fig:sna_development}(a)] with the course seating chart (provided by the course instructor) confirms that students mostly reported ties with their group members sitting at the same table. This suggests that this collection occurred too early in the semester for meaningful classroom connections to have formed and the reported interactions occurred mostly out of convenience stemming from physical proximity.

In the second collection, outdegree becomes a significant predictor of final grade, even when controlling for GPA. The outdegree model ($R^2_{adj}=0.296$) explains about $12\,\%$ more of the variance than GPA alone ($R^2_{adj}=0.175$). The LRT further confirms that the full model with outdegree is significantly better than the benchmark. It is interesting that outdegree remains significantly correlated with grades, even when controlling for GPA, but for indegree even the simple model is not. This is to say that a student's self-reported social interactions matter, but not interactions reported by their peers. Such a result indicates the importance of a student's engagement beliefs, i.e., a student's own perception of their engagement. 

\begin{table}[t]
\renewcommand{\arraystretch}{1.1}
\renewcommand{\tabcolsep}{4pt}
\caption{Linear regression results for the full models (final grade $\sim$ GPA + centrality) for centralities with at least one occurrence of significant correlations in simple models. Reported $p$ values have been Bonferroni adjusted at the collection level. Significant $p$ values are marked with the appropriate number of asterisks. In all reported cases, the $p$ value of $F$ statistics is $p<0.001$. Models in which centrality is not a significant predictor and models that failed the bootstrap test are omitted for clarity. This includes the entire week 2.} \centering
\begin{tabular}{lrcccc}
\hline \hline
Centrality &   &\multicolumn{4}{c}{Regression statistics}\\
& & Week 6 & Week 8 & Week 11 & Week 13 \\
\hline
Indegree &  $\beta_{GPA}$ & - & - & $0.42^{***}$ & $0.41^{**}$ \\
& $\beta_{I}$ & - & - & $0.52^{***}$ & $0.30^{*}$ \\
& $F_{2,70}$ & - &  - & 32.3 & 14.3 \\
& $R^2_{adj}$  & -  & - & 0.465 & 0.272 \\ 
\hline

Outdegree & $\beta_{GPA}$ & $0.37^{**}$ & $0.34^{**}$ & $0.44^{***}$ & $0.40^{**}$ \\
& $\beta_{O}$ & $0.34^{*}$ & $0.43^{***}$ & $0.38^{**}$ & $0.37^{**}$ \\
& $F_{2,70}$ & 16.1 & 21.3 & 19.3 & 18.3 \\
& $R^2_{adj}$ & 0.296  & 0.361 & 0.337 & 0.324 \\ 
\hline

Closeness & $\beta_{GPA}$ & - & $0.34^{*}$ & $0.43^{***}$ & $0.41^{***}$ \\
& $\beta_{C}$ & - & $0.41^{**}$ & $0.53^{***}$ & $0.39^{**}$  \\
& $F_{2,70}$ & - & 19.5 & 33.0 & 19.3 \\
& $R^2_{adj}$ & - & 0.339 & 0.471 & 0.337 \\

\hline \hline
\multicolumn{6}{l}{\footnotesize ***$p<0.001$, **$p<0.01$, *$p<0.05$}
\label{tab:full_models}
\end{tabular}
\end{table}

\begin{table}[b]
\renewcommand{\arraystretch}{1.1}
\renewcommand{\tabcolsep}{8pt}
\caption{Summary table of the likelihood ratio test (LRT) comparing full models (final grade $\sim$ GPA + centrality) to the base model (final grade $\sim$ GPA). Reported $p$ values have been Bonferroni adjusted based on the total number of LRTs (i.e., scaling by a factor of nine). Significant $p$ values are marked with the appropriate number of asterisks. Nonsignificant values have been omitted for clarity. This includes the entire week 2.} \centering
\begin{tabular}{lcccc}
\hline \hline
Centrality & \multicolumn{4}{c}{$\chi^2(d.o.f.=1)$}\\
&  Week 6 & Week 8 & Week 11 & Week 13  \\
\hline
Indegree & - & - & $30.8^{***}$ & $8.3^{*}$ \\ 
Outdegree  & $10.8^{**}$ & $17.8^{***}$ & $15.1^{***}$ & $13.7^{**}$ \\
Closeness & - & $15.3^{***}$ & $31.6^{***}$ & $15.1^{***}$ \\ 
\hline \hline
\multicolumn{5}{l}{\footnotesize ***$p<0.001$, **$p<0.01$, *$p<0.05$}
\label{tab:LRT}
\end{tabular}
\end{table}

Results from the following three collections are somewhat similar to each other. In particular, we find that in addition to outdegree, closeness becomes significantly correlated with the course grade when controlling for GPA starting at the third collection and indegree starting at the fourth collection. The LRT further confirms that the corresponding full models are significantly more informative than the benchmark model [relation~(\ref{eq:benchmark})]. Furthermore, the $R^2_{adj}$ values reveal that all these models predict future performance better than the benchmark does, explaining up to $47\,\%$ of the variance in course grade, compared to the benchmark's explanatory power of only about $20\,\%$. Interpreting these results in terms of the classroom, the significance of outdegree and indegree confirms the importance of direct interaction with peers. The significance of closeness indicates the importance of integration into the whole network in a broader sense.

\begin{figure}[t]
\centering
 \includegraphics[width=0.49\textwidth]{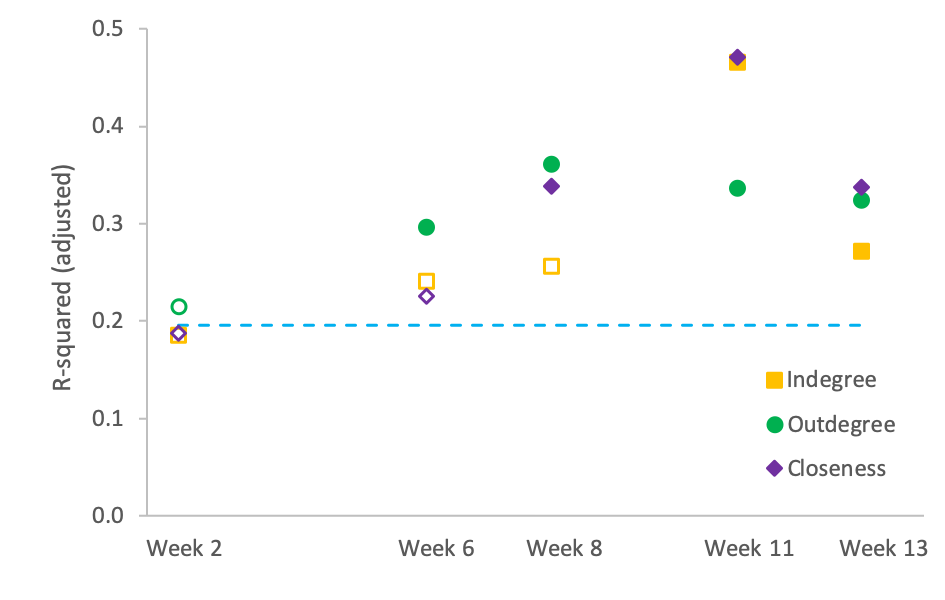}
 \caption{Scatter plot of $R^2_{adj}$ values for the full models (final grade $\sim$ GPA + centrality) for indegree (yellow square), outdegree (green circle), and closeness (purple diamond), shown chronologically. The horizontal dashed blue line shows  $R^2_{adj}$ for the base model (final grade $\sim$ GPA) for comparison. We observe a general trend of increasing predictive power as time passes. Markers for $R^2_{adj}$ for models that were not statistically significant are left empty.}
\label{fig:Rsq_plot}
\end{figure}

In the findings described above, we observe a general trend for the $R^2_{adj}$ values to increase over time, with peaks at collection 4 (week 11) for indegree and closeness, and at collection 3 (week 8) for outdegree (see Fig.~\ref{fig:Rsq_plot}). The explanatory power of models containing mid- or late-semester student interactions is up to 2.3 times greater than the explanatory power of GPA alone. 

%%%%%%%%%%%%%%%%%%%%%%%%%%%%%%%%%%%%%%%%%%%%%%%%%%%%%%%%%%%%%%
\subsection{Which centralities matter?}
%%%%%%%%%%%%%%%%%%%%%%%%%%%%%%%%%%%%%%%%%%%%%%
Of the four tested centrality measures, closeness is most strongly correlated with final grades. It also seems to be the best representation of Tinto's student integration model. This is because closeness represents a student's ties to the classroom as a whole---granting easy access to information as well as academic and social support from a robust group of peers without the need to go through many intermediaries. It thus seems to be the most valuable centrality measure to study when considering questions of student engagement and persistence~\cite{Zwolak17-NIP, Zwolak18-PIN}. Another point in favor of closeness is that it explains the most variance of all measures in its peak at week 11 with $R^2_{adj}$=0.47. Therefore a midsemester closeness measurement represents the best way we discovered to predict the end-of-semester final grade.

The outdegree model peaks at week 8, and is also the best predictive model available at that time, with $R^2_{adj}$=0.361. Furthermore, the outdegree model's predictive power emerges the earliest, at week 6, with $R^2_{adj}$= 0.296 (1.5 times better than the GPA model alone). As such, it is the best choice for predicting performance early on in the semester. Indegree, on the other hand, while significant for the last three collections when looked at alone, looses its predictive power when GPA is considered until week 11.

Finally, betweenness is never a significant predictor, even when GPA is not included. This result is consistent with our theoretical framework, as betweenness represents a very specific type of position in the network characterized by being a bridge between otherwise disparate groups. Such a position may be indicative of an ancillary status as a member of multiple small groups, rather than being well-connected within any one group [compare the bridging position of the orange nodes in the network from the beginning of semester shown in Fig.~\ref{fig:sna_development}(a) and their peripheral position by week 11 shown in Fig.~\ref{fig:sna_development}(b)]. The active engagement nature of the MI curriculum does not seem to provide an environment where such bridging positions would be easy to establish nor benefit students. Our study suggests that this type of engagement, while important to prevent network fragmentation, does not serve the interest of supporting a student's own academic performance.  

%%%%%%%%%%%%%%%%%%%%%%%%%%%%%%%%%%%%%%%%%%%%%%%%%%%%%%%%%%%%%%
\subsection{Community formation over time}
%%%%%%%%%%%%%%%%%%%%%%%%%%%%%%%%%%%%%%%%%%%%%%
We observe that, as the semester progresses, interpersonal interactions become more significant. At the local level this is indicated by outdegree; at the global level by closeness. On the first collection, none of the measures are significantly correlated with grades and on the second only one local level measure is significant (outdegree). On the following three collections, three out of the four considered measures, i.e., indegree, outdegree, and closeness, are significantly correlated with final grades, with outdegree and closeness remaining significant even when controlling for GPA. This indicates that in the first half of the semester there is little to no effective integration occurring; the little integration that does occur exists only with nearest neighbors and is predicated on each student's self-perception of their own behavior. Yet, by midsemester,  interactions at the local and global scales all predict higher performance. Such a change indicates a time development of the classroom community, wherein the student interactions in the second half of the semester effectively predict final grade at the end of the course. 

\begin{figure}[t]
\centering
 \includegraphics[width=0.48\textwidth]{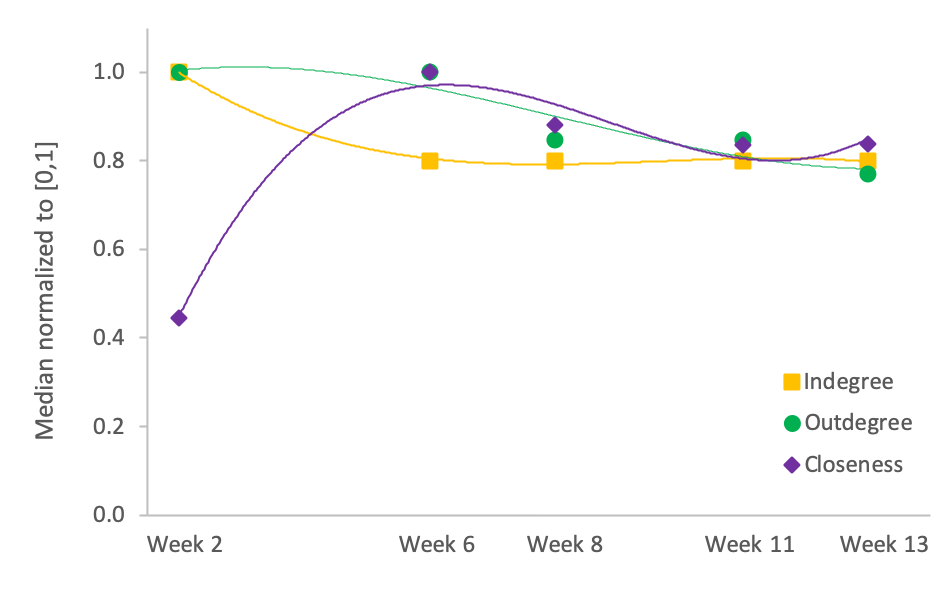}
 \caption{Scatter plot of the centrality median change (with polynomial trend lines) for indegree (yellow square), outdegree (green circle), and closeness (purple diamond), shown chronologically. To compare the overall trend, the median values are normalized to $[0,1]$ for each centrality. We observe a general trend of decreasing median value for all presented measures, with equilibrium reached around week 8.}
\label{fig:median_plot}
\end{figure}

%%%%%%%%%%%%%%%%%%%%%%%%%%%%%%%%%%%%%%%%%%%%%%
\subsubsection{Group dynamics}
Time development is also apparent from viewing the network diagrams of the classroom as in Fig.~\ref{fig:sna_development}. Early in the course [Fig.~\ref{fig:sna_development}(a)], the majority of ties exist between same-group members and the six-person seating arrangement is readily apparent. Later in the course, seating groups are almost completely indistinguishable in favor of a more unified network, indicating classroom-wide integration [Fig.~\ref{fig:sna_development}(b)]. 

Considering both structured and unstructured cross-group interactions, the development of the network over time may be explained by a number of factors. For example, students may be developing a rapport as they get to know each other better. As they change groups and are forced to work with new assigned seatmates, they might recall better working relationships with former seatmates and seek them out instead of (or in addition to) working with the new group members. Growing rapport would also explain increased cross-group interactions in board meetings: students who have previously worked together well will likely communicate more meaningfully in large discussions than if they had not. As the semester progresses, increasing number of students in a given board meeting have prior work experience with the other students in attendance. This interpretation is corroborated by anecdotal observations from the course instructors (and this study's authors on data collection days) and by follow-up interviews with students, which indicated that students did indeed become more comfortable talking to each other in the board meetings as the semester progressed. This time development may also indicate a shift in students' perception of the word meaningful from working together on a problem to discussing phenomena in board meetings. In fact, evidence from an in-progress study suggests that students who have taken a first semester of MI have more positive attitudes toward MI in their second semester and more quickly embed themselves in the classroom network.

%%%%%%%%%%%%%%%%%%%%%%%%%%%%%%%%%%%%%%%%%%%%%%
\subsubsection{Shift in perception: Towards efficient networks}
We find that students' engagement changes in a nonintuitive way throughout the semester. Although we hypothesized that over time the network density will increase, we found that the number of reported interaction actually slightly decreased, from 358 on the first collection to 302 by the end of the semester (with $N_{Week\;2}=359$, $N_{Week\;3}=302$, and $N_{Week\;4}=312$; (see Table~S2 the Supplemental Material~\cite{supp} for comparison of network characteristics). The median values of indegree and outdegree also decrease over time (see Fig.~\ref{fig:median_plot}), with indegree reaching equilibrium around week 8 and outdegree around week 6. Closeness median oscillates over the first half of the semester, reaching its peak at week 6 and stabilizing by week 8. The average values for all three measures, renormalized to account for the number of students whose names appeared on a given collection, as well as the raw number of ties per student also decreases over the first half of the semester before stabilizing around week 8. This assures us that the observed time development of the in-class network is not simply a measurement or reporting error (i.e., that students know only a few of each others' names at the beginning of the semester but learn and report more names over time). 

This unexpected evolution of the in-class network does not mean that engagement is stagnant or decreasing; while the \textit{number} of ties trends slightly down, the structure of the network also changes dramatically. What we see in the network evolution is a shift in the distribution of ties from primarily seatmate connections to meaningful cross-group connections. This shift in the connectedness is clearly visible when comparing Figs.~\ref{fig:sna_development}(a) and ~\ref{fig:sna_development}(b): While there are more ties overall and more ties per student reported on the first collection [shown in Fig.~\ref{fig:sna_development}(a)], the network is more scattered and has more weakly interconnected clusters than the network from the fourth collection [presented in Fig.~\ref{fig:sna_development}(b)]. We interpret this change as evidence of a {\it selection effect}: While the quantity of interactions decreases over time, their quality improves, suggesting that, after the initial exploratory in nature interactions, students begin to make more calculated decisions regarding whom to interact with. During the first half of the semester, they cast a wide net to interact with each other but, as they find that not all of these interactions are desirable, by midsemester they begin to interact with slightly fewer but more selectively chosen people. It may be that students need enough time to find the ``right'' people with whom to collaborate with and who can support their academic development. The fact that the total number of interactions reported by all students on each collection and the average number of ties reported per student present does not increase over time further supports the selection effect.

%%%%%%%%%%%%%%%%%%%%%%%%%%%%%%%%%%%%%%%%%%%%%%%%%%%%%%%%%%%%%%
\section{Conclusions}\label{sec:conclusion}
%%%%%%%%%%%%%%%%%%%%%%%%%%%%%%%%%%%%%%%%%%%%%%%%%%%%%%%%%%%%%%
In this study, we establish a relationship between engagement, as measured by network centrality, and future academic performance. Examination of the in-class community development over the course of the semester allows us to not only identify {\it which} centrality measures are useful for capturing the important aspects of student engagement but also to determine {\it when} community interactions begin to be predictive of performance. In the process, we find evidence that the relationship between engagement and performance is more nuanced than we expected. While our study has some limitations, it inspires us to look toward future lines of research. 

%%%%%%%%%%%%%%%%%%%%%%%%%%%%%%%%%%%%%%%%%%%%%%%%%%%%%%%%%%%%%%
\subsection{Engagement predicts academic performance}
%%%%%%%%%%%%%%%%%%%%%%%%%%%%%%%%%%%%%%%%%%%%%%
We find three out of four centrality measures, acting as proxies for student engagement, to be significantly correlated with future academic performance. Models based on these measures predict future performance even when controlling for past performance --- and better than past performance alone. 

The best predictive power of student engagement comes from the closeness model ($final \; grade \sim GPA + closeness$) at fourth collection. This result is powerful for two reasons. From a theoretical perspective, closeness centrality represents how easily a student may interact with all members of the classroom community using as few intermediaries as possible. Thus, it is the closest analog to Tinto's conception of student integration. From a data-driven perspective, closeness yields the most powerful model in the entire study, explaining nearly half of all the variance in students' course grades. Therefore, measuring closeness centrality around week 11 (approximately two-thirds of the way into the semester) seems to be the best way to capture student engagement as relevant to final course performance.

It is also useful to consider the earliest predictive power of student engagement. The outdegree model's predictive power peaks as early as the second collection (week 6), explaining nearly $30\,\%$ of the variance in final grade. By the third collection (week 8), the model with outdegree and GPA improves to account for over $36\,\%$ of the variance, which is only 10 percentage points less than the best model and is the third-best model overall. The fact that students' self-perceptions of engagement reaches peak predictive power so early is empowering to course instructors. Since predictions about students' course grades are accessible as early as week 8, there is still plenty of time for instructors to enact interventions that promote student engagement,  thus positioning attrition-risk students for increased success.

%%%%%%%%%%%%%%%%%%%%%%%%%%%%%%%%%%%%%%%%%%%%%%%%%%%%%%%%%%%%%%
\subsection{Network evolution}
%%%%%%%%%%%%%%%%%%%%%%%%%%%%%%%%%%%%%%%%%%%%%%
We expected the number of network ties to increase over time as more time affords students more opportunities for engagement. Instead, we find the total number of ties as well as the number of ties per person to decrease slightly over time, while at the same time the predictive power of the models increases. This suggests that, rather than having more interactions over time, students change their interaction patterns to best accommodate their social and academic needs (the {\it selection effect}). 

This change can be interpreted in several ways. It could be that students need time to find the right people with whom they can effectively collaborate; it may also be that they need time to get used to the concept of a collaborative-learning physics classroom and see each other as valuable learning resources. Another possible interpretation is that students' understanding of meaningful interaction changes over time, perhaps due to an evolving comprehension of which peer interactions are important to their classroom experience (social, academic, structured, unstructured). A slight shift in weight distribution, from $p_1=4.5\,\%$, $p_2=25.1\,\%$, and $p_3=70.4\,\%$ during the first collection (where $p_i$ denotes the percentage of ties with weight $i$) to $p_1=5.6\,\%$, $p_2=34.1\,\%$, and $p_3=60.3\,\%$ during the last collection, supports the latter hypothesis. However, it could also simply be the case that the second half of the semester is more important: whether a student begins the semester strongly or weakly, if they work hard and do well in the middle and towards the end of the semester, they will earn a high final grade purely due to the algebra of how course grades are calculated. We acknowledge that these interpretations are speculative---network analysis alone is inadequate to understand how students conceive of their network engagement. Qualitative follow-up is necessary to provide a more detailed understanding.

%%%%%%%%%%%%%%%%%%%%%%%%%%%%%%%%%%%%%%%%%%%%%%%%%%%%%%%%%%%%%%
\subsection{Practical applications for network methods}
%%%%%%%%%%%%%%%%%%%%%%%%%%%%%%%%%%%%%%%%%%%%%%
The results from this study have practical applications for the use of network analysis in an academic setting, especially the classroom. We find significant predictive power primarily between weeks 8 and 13 of a 16-week course, and the results are fairly consistent. This suggests that network data can be collected only once during this time frame without major information loss. Such streamlining of the data collection process will benefit researchers who can collect and analyze less data; student participants will face less survey fatigue; and instructors will not have to disrupt their class for multiple data collections. Minimal data requirements ensure a lower barrier to implementation.

Moreover, such insight provides markers for universities seeking to retain students through careers in science that require courses like introductory physics. Failure to pass introductory science courses the first time around exacts a price from both students and the institutions that keep them from persisting in their majors at greater rates \cite{BreweDou2018, eagan2008closing}. Universities with high populations of students from minoritized backgrounds, such as FIU, should take special note of the practical applications of our results as a form of gauging student embeddedness with the express purpose of facilitating success and/or providing targeted support to students. Both Nora and Tinto remind us of the value of social embeddedness as a characteristic of students who persist in higher education \cite{Nora03-AHH, Tinto06-RPWN}.

%%%%%%%%%%%%%%%%%%%%%%%%%%%%%%%%%%%%%%%%%%%%%%%%%%%%%%%%%%%%%%
\subsection{Limitations and future work}
%%%%%%%%%%%%%%%%%%%%%%%%%%%%%%%%%%%%%%%%%%%%%%
There are some limitations to the results of this study. First, the sample we look at represents only one section of an introductory physics course ($N=73$ students) offered at FIU. Further investigation of more courses, at both introductory and upper-division level, and more varied populations should be studied. Second, our interpretations of why academic performance is predicted by certain centrality measures at certain times (and not other measures at other times) are speculative. Although we have offered several possible interpretations grounded in a framework of engagement theories, qualitative work must be done to determine which of these interpretations (if any) represents the mechanism(s) underpinning our results.

Future work should further explore the reciprocal relationship between performance and engagement. Structural equation modeling may be used to disentangle the direct effect of past performance on future performance from the indirect effect of past performance on engagement, thus influencing future performance. Accounting for interactions occurring outside of the classroom in a more casual setting to determine whether they are related to in-class interactions, performance, and other outcomes, would also be valuable. 

The ties' weights can be utilized to capture different information. While this study uses them to quantify the frequency of interactions, one could instead use them to assess their quality by explicitly asking students to, e.g., rank their meaningful interactions on a scale. Students could also be asked to distinguish among various kinds of interactions (e.g., friendship vs content related) to determine if different peer interactions are related to different outcomes. 

Finally, qualitative methods (especially interviews) should be used as a follow-up to investigate how students perceive their engagement in learning communities---both inside and outside the classroom. For example, the literature around sense of belonging (see Ref.~\cite{OKeeffe2013}) supports the notion that students who feel they belong to a community of peers are more likely to be retained in courses and persist through their academic careers. While it is plausible that students who feel a sense of belonging in the classroom have high closeness centrality (or vice versa), qualitative studies could help confirm the connection between the quantitative measure of closeness to the affective construct of belongingness.

%%%%%%%%%%%%%%%%%%%%%%%%%%%%%%%%%%%%%%%%%%%%
\begin{acknowledgments}
Supported by NSF PHY 1344247. We would like to thank the course instructors for facilitating data collection. All protocols in the project were approved by the Florida International University Institutional Review Board (IRB-13-0240 exempt, category 2).
\end{acknowledgments}

%%%%%%%%%%%%%%%%%%%%%%%%%%%%%%%%%%%%%%%%%%%%
\bibliographystyle{plain}

%%%%%%%%%%%%%%%%%%%%%%%%%%%%%%%%%%%%%%%%%%%%
\end{document}